# Complete mapping of interacting charging states in single coupled colloidal quantum dot molecules

*Yossef E. Panfil, Jiabin Cui[†], Somnath Koley & Uri Banin\**

Institute of Chemistry and the center for nanoscience and nanotechnology, The Hebrew University of Jerusalem, Jerusalem 91904, Israel.

\* Corresponding author. Email: uri.banin@mail.huji.ac.il (U.B.)

**Abstract**: Colloidal Quantum Dots (CQDs), major building blocks in modern opto-electronic devices, have so far been synthesized with only one emission center where the exciton resides. Recent development of coupled Colloidal Quantum Dots Molecules (CQDM), where two core-shell CQDs are fused to form two emission centers in close proximity, allows to explore how charge carriers in one CQD affect the charge carriers in the other CQD. Cryogenic single particle spectroscopy reveals that while CQD monomers manifest a simple emission spectrum comprising a main emission peak with well-defined phonon sidebands, CQDMs exhibit a complex spectrum with multiple peaks that are not all spaced according to the known phonon frequencies. Based on complementary emission polarization and time-resolved analysis, this is assigned to fluorescence of the two coupled emission centers. Moreover, the complex peak structure shows correlated spectral diffusion indicative of the coupling between the two emission centers. Utilizing Schrodinger-Poisson self-consistent calculations, we directly map the spectral behavior, alternating between neutral and charged states of the CQDM. Spectral shifts related to electrostatic interaction between a charged emission center and the second emission center are thus fully mapped. Furthermore, effects of moving surface charges are identified, whereby the emission center proximal to the charge shows larger shifts. Instances where the two emission centers are negatively charged simultaneously are also identified. Such detailed mapping of charging states is enabled by the coupling within the CQDM and its anisotropic structure. This understanding of the coupling interactions is a progress towards quantum technology and sensing applications based on CQDMs.

**Introduction**

Colloidal Quantum Dots (CQDs) have reached a high level of control and are nowadays bright and stable emitters.[1–3] Numerous hetero-structures have been developed including core-shell spheres,[4,5] dot-in-rod,[6,7] rod-in-rod,[8] core-crown and core-shell nanoplatlets[9,10] and nano-dumbelles.[11,12] Accompanied by the wet-chemistry flexible manipulation of the nanocrystals, this has led to numerous applications. A lead example where CQDs are already commercialized is their use in information displays,[2,3,13] showcasing the feasibility of their scalable production and their robustness. Further applications of CQDs ranging from lasers,[14] solar cells[15] and 3D printing might appear in the future.[16,17] However, so far, CQDs have been mostly synthesized with one emission center, where the electron-hole pair created upon excitation are cooling down into its band-edge. Although more complicated heterostructures with two emission centers have been synthesized,[18,19] the two emission centers, where the charge carrier resides, were typically not close enough to generate electrostatic coupling between the charge carriers.

Herein, we perform a cryogenic spectroscopic study on single colloidal quantum dot molecules (CQDMs) accompanied by theoretical modelling revealing their intricate emission spectrum and its dynamical behavior. Such CQDMs are composed of two fused core-shell CdSe-CdS QDs in close proximity.[20–22] It is important to recall that the famous DiVincenzo criteria for scalable quantum computing platform was originally proposed on QD molecules.[23] It states five criteria for a quantum computing platform, which are all naturally present in quantum dot molecules. So far, the major progress in realization of Qubits gates was achieved in either gate defined QD molecules[24,25] or for epitaxial self-assembled QD molecules.[26] However, CQDMs are typically much smaller in size and in closer proximity, and therefore can facilitate stronger coupling with larger energy level separation important also for higher temperature applications. Furthermore, they are prepared with wet-chemistry methods, which are scalable and applicable for flexible patterning and fabrication of devices in the emergent field of quantum technologies.

Single particle spectroscopy is particularly well suited for this study, as was demonstrated in its utilization for deciphering charging states in colloidal QDs from the intensity time trace and lifetime dynamics or from its emission spectrum at cryogenic temperatures.[27–31] The orientation of the nanocrystal can also be obtained from its emission polarization characteristics.[32] However, these measurements were not performed all at once. This becomes important especially for the exploration of CQDMs, which manifest two emission centers. Each one of the QDs comprising the CQDM might be neutral, charged separately or charged simultaneously along with the other

QD. In addition, since the two QDs might be in the same or different orientations,[20] their emission polarization angle, which is dictated by the Wurtzite c axis in the case of the CdSe-CdS core/shell nanocrystal system, will vary accordingly and therefore can shed light on their relative alignment and fusion configuration within the CQDM. Moreover, because of the limited number of photons which are emitted out of a single nanocrystal before it bleaches, it is essential to extract all the possible information simultaneously.

Through the combination of the simultaneous cryogenic spectral, lifetime and polarization measurements, we show on a single particle level that the two QDs comprising the CQDM are affecting each other. Firstly, when one of the QDs is becoming charged, this can be identified via the change in its various characteristics such as a spectral shift accompanied by lower intensity and shorter lifetime. This is seen to slightly change the spectrum of the other QD via electrostatic interaction. In some instances of a change in the charging state on one QD, a large effect is observed on the other QD. Employing numerical simulations, we attribute this to surface charge movement from the charged QD to the surface of the other one. Notably, we also observe instances where the two QDs are becoming negatively charged simultaneously. This work therefore unravels in detail, through a "detective-like" approach, the origin of each and every peak in the complex CQDM emission spectrum revealing various coupled charged states, the transitions between them, and the effects of surface charge.

The implications of such detailed mapping and understanding of the emission and charging effects in CQDM, is also in it being a step towards their improved control. It provides understanding of how to instill desired emissive properties in these systems, while serving as a basis for considering various innovative electro-optic devices and quantum technology applications utilizing such CQDMs.[33–36]

**Results and Discussion**

**CQDM samples and the Single Particle Cryogenic Optical Microscope Setup**

CdSe/CdS core shell nanocrystal were synthesized and then linked, fused and purified according to a recent protocol.[20] Briefly, silica spheres with diameter of ~200 nm were used as a template for the controlled dimers formation. The first monomer layer is bound to the thiol functionalized silica sphere surface, followed by masking of the surface with a thin silica layer to fixate the monomers and block further binding to the silica itself. The bound CQDs are functionalized by tetra-thiol molecules serving as a linker. Then the solution is exposed to the second CQD layer, which selectively bind to the first CQDs through the thiol linkers. In the next step, the silica spheres are etched selectively by HF, and the solution is cleaned from silica fragments. The solution is next taken to a fusion step at moderate temperature, which forms a continuous crystalline connection between the two CQDs thus forming the CQDM. Purification by size-selective separation is then applied to separate out monomers and higher order oligomers, while achieving a fraction enriched in dimers.

In the studied sample we aim at CQDM with stronger coupling achieved for a CdSe core radius of 1.3nm and the CdS shell thickness of 2.1nm (Supporting Information fig. S11 shows the monomers size dispersion). Figure 1a shows a transmission electron microscopy (TEM) image demonstrating the formation of quantum dot molecules. The high angle annular dark field (HAADF) scanning transmission electron microscopy (STEM) characterization shows the fusion of the two CdSe/CdS nanocrystals forming the CQDMs (Fig. 1b). The core-shell architecture in the CQDMs was maintained as demonstrated by the energy dispersive X-Ray spectroscopy (EDS) measurement on the same particle (Fig. 1c). A continuous distribution of cadmium (both in core and shell) is identified throughout the projection of the CQDM. However, selective regions of the selenium (only in core) are clearly identified signifying the cores locations.

For the cryogenic single particle spectroscopy studies, the CQDMs were dispersed by spin casting a PMMA and toluene solution with low CQDM concentration, on a silicon substrate which was mounted inside a 4K cryogenic single particle microscope/spectroscopy system (Figure 1d). A 475nm pulsed laser (50 psec pulses, 1 MHz, typically intensity of 300 nW), focused by a cooled objective lens (numerical aperture of 0.82) on the sample plane, was used to excite single CQDMs. The emission of the single CQDMs is then collected using the same objective and filtered from the excitation light by a Dichroic Mirror (DM) and then split by a 50:50 Beam Splitter (BS) - half into an Avalanche Photo Diode (APD) and half into a spectrograph through a rotating $\frac{\lambda}{2}$ wave-plate (rotating 5º every ~10 sec) and Polarization Beam Splitter (PBS) which displace the vertical (V) and horizontal (H) components spatially into separate beams. The two emission spectral components are then imaged on an EMCCD (400 frames with exposure time of 0.5 sec). Using a Time Tagger, triggered either upon excitation of the laser, upon $\frac{\lambda}{2}$ wave-plate rotation, upon photon detection in the APD or upon exposure of the EMCCD, we are able, by post processing, to trace each and every detected photon to the relevant respective measurement condition (Fig. 1d). The typical overall measurement time from a single particle is 200 sec.

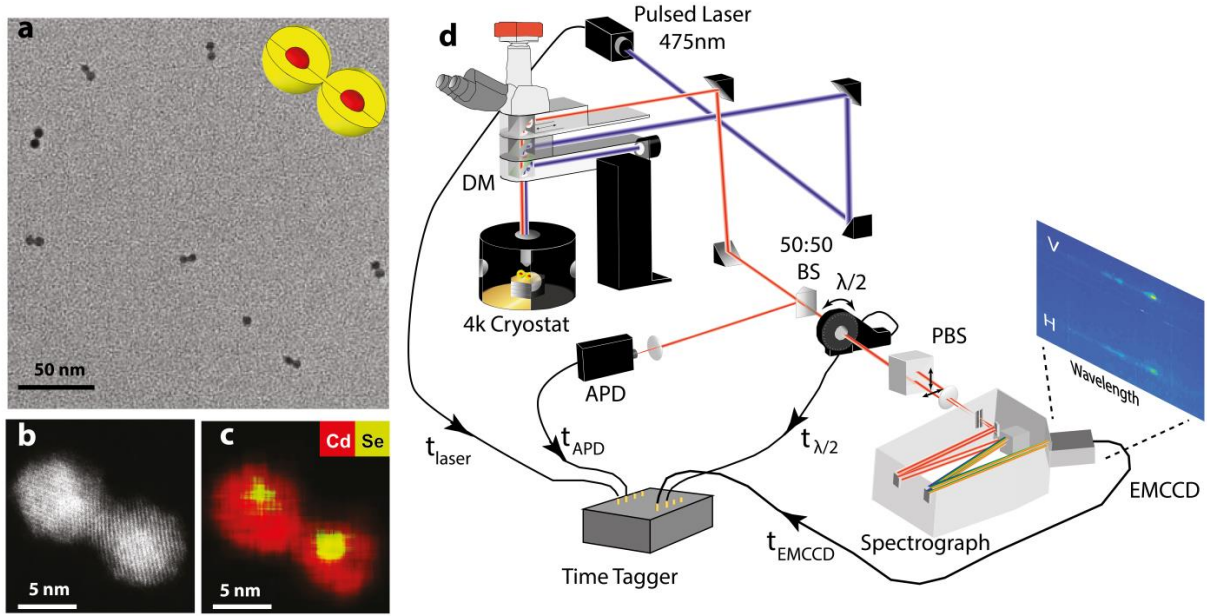

**Figure 1. CQDMs characterization and single particle optical microscope setup.** (a) TEM image of a grid containing 1.3nm/2.1nm CdSe/CdS core radius/shell thickness CQDMs. (b) HAADF STEM characterization of a single CdSe/CdS CQDM. (c) EDS measurement on the same particle showing the distribution of cadmium and Selenium throughout the CQDM. (d) Single particle optical microscope comprised of a 4K cryostat, 475nm pulsed laser, Dichroic Mirror (DM), Beam Splitter (BS), APD, rotating $\frac{\lambda}{2}$ wave-plate, Polarization Beam Splitter (PBS), spectrograph, EMCCD and a Time Tagger. The inset shows one frame with the two replicas of the spectrum, one for the Vertical (V) and one for the Horizontal (H) components of the emission linear polarization collected by the EMCCD.

**Single CQD vs. single CQDM spectrum & lifetime data.**

Figure 2 presents data taken from a single CQD (a-h) and from a single CQDM (i-p). Figure 2b presents 400 consecutive spectra (0.5 sec each frame, the vertical and horizontal polarization replicas are added together). Starting from the simpler CQD spectra, typically it may show blinking and spectral diffusion, and therefore spectral clustering is performed where similar regions are combined and analyzed together. The CQD spectrum manifests two major clusters (indicated by colored stripes at the top and bottom of the time dependent spectrum presented in figure 2b). Firstly, the blue cluster for which figure 2g presents the accumulated spectrum. The major peak in the blue cluster-spectrum is the narrow Zero Phonon Line (ZPL) emission from the band edge of the nanocrystal, while the smaller peaks, red shifted by 27meV and 37meV, are the CdSe and CdS longitudinal optical (LO) phonon replicas, respectively. The lifetime data (fig. 2e), which was generated only by photons arriving during the blue cluster, contains a short component (~1nsec) and a long component (~113nsec), both typical to neutral CQDs at cryogenic temperatures.[37] The short component relates to carrier cooling and the long component to the radiative recombination from the lowest ('dark') state. The intensity time trace taken simultaneously from the APD (Fig. 2c), also proves that the blue cluster is the neutral 'on' state of the CQD according to the high intensity counts during this time.

However, the red cluster in figure 2b, for which the accumulated spectrum is presented in figure 2h, manifests weaker emission (fig. 2c), and the ZPL spectrum is red shifted (fig. 2h) by ~15meV relative to the blue cluster corresponding to the neutral CQD. These signatures correspond to a negatively charged CQD. For positive charge it would have been blue shifted as calculated using our numerical simulation. In addition, Previous studies have suggested that photoionization produces primarily negatively charged species.[38,39] Moreover, because of the quasi-type-II nature of CdSe/CdS core shell QDs, the positive trion should have lower QY compared to the negative trion, resulting in a dark state.[40] For a charged CQD, the non-radiative Auger decay dominates and accordingly, the lifetime contains only a short component of ~6nsec[41] (fig. 2d). This has two origins, first - non-radiative Auger relaxation in such case would lead to fast decay. Second – the presence of the extra carrier should lead to bright state occupation at the band edge. [38] These two clusters are well identified in the Fluorescence Lifetime Intensity Distribution (FLID) presented in figure 2f, which contains one area with low intensity and short lifetime (from the red cluster), and one area of high intensity and high counts (from the blue cluster).

To these measurements we also added the emission polarization analysis of the blue and red cluster's ZPL peaks, $p(\theta) = \frac{I_V(\theta) - I_H(\theta)}{I_V(\theta) + I_H(\theta)}$, and then fitting it to $p(\theta) = p \cdot cos(4\theta - 2\varphi)$ where $\theta$ is the rotation angle of the $\frac{\lambda}{2}$ wave-plate and $\varphi$ is the angle of polarization with respect to some reference angle. The emission polarization angle is not changing upon charging, as expected (fig. 2a). This typical CQD spectra manifests characteristics consistent with earlier studies, providing strong validation to our measurement setup, the methodology of the cluster multi-characteristics analysis, and to the quality of the CQD monomers forming the CQDM.

Focusing next on the CQDM spectrum, a much more elaborate behavior is seen. The 400 consecutive spectra contain more peaks (fig. 2j). Addressing the origin of the increased number of peaks, polarization analysis of the two major peaks, the ones in the magenta and orange regions of the spectrum (sides of Fig. 2j), yields two significantly different emission polarization angles (fig. 2i), unlike the case of the monomer CQD discussed above. The two polarization angles from the different peaks teaches us that the emission may arise from either emission centers in the CQDM. This arises as the excitation can lead to branching of the excitons relaxing to either emission center with some probability. Since the emission polarization angle is dictated by the Wurtzite c axis - the two QDs, marked as QD1 and QD2, are with different orientation.[32,42] Such different orientations indicate that the particular CQDM manifests an heteronymous plane attachment in which the two QDs are fused trough different facets.[20]

Clustering the CQDM spectral bins yielded three major clusters, further establishing the role of two emission centers and charging effects. The accumulated spectrum of the blue cluster is plotted in figure 2p (blue) and manifests two major peaks with nearly identical intensities, separated by 20 meV. The lifetime data (fig. 2n) contains short (~2nsec) and long components (~70nsec) typical to the neutral CQDS at cryogenic temperatures. Since the short component in the lifetime cannot be solely attributed to one of the peaks, this means that the two major peaks are the ZPLs arising from each one of the QDs comprising the CQDM emitting from their neutral state. The spectral shift between QD1 and QD2 is attributed to a change in the confinement energy related to their different size.

The red cluster shows low intensity (fig. 2k), a short lifetime (fig. 2m, ~1 & 9nsec components), and red shifted emission (fig. 2j&p) mostly from QD2. We infer that in this cluster, QD2 is negatively charged and QD1 is hardly emitting, probably due to a trap on the surface. The green cluster is also manifesting low intensity, but the lifetime data shows a long component (fig. 2l, ~70nsec) and emission dominantly from QD1 that is at the same energy as in the blue cluster (fig. 2p). We infer that in this case, QD1 is neutral while QD2 is not emissive, likely due to a trap on its surface. It should be noted that although QD1 has higher energy than QD2, we don't observe any FRET process. An evidence for FRET would be, for example, if the lifetime taken from QD1, the blue-shifted line (while this line is strong and the red-shifted line is dimmed as in the green cluster), contain only short component or more pronounced short component in the lifetime due to FRET. However, this is not the case (comparing fig. 2l to fig. 2n). FRET is greatly dependent on the spectral overlap between the donor and acceptor. In low temperatures the spectral lines are very narrow prohibiting any spectral overlap. This means that FRET is not taking any role at low T, Nevertheless, FRET can play a major role in room temperature.

One should notice that the intensity of the red and green clusters are nearly the same (fig. 2k). This is also seen from the FLID data (fig. 2o) where there are no distinct states in the intensity and lifetime of CQDMs. However, this analysis shows that the state of the CQDM is different. This further signifies the importance of such analysis instead of the commonly used intensity clustering,[27,41] in resolving the state of the CQDMs.

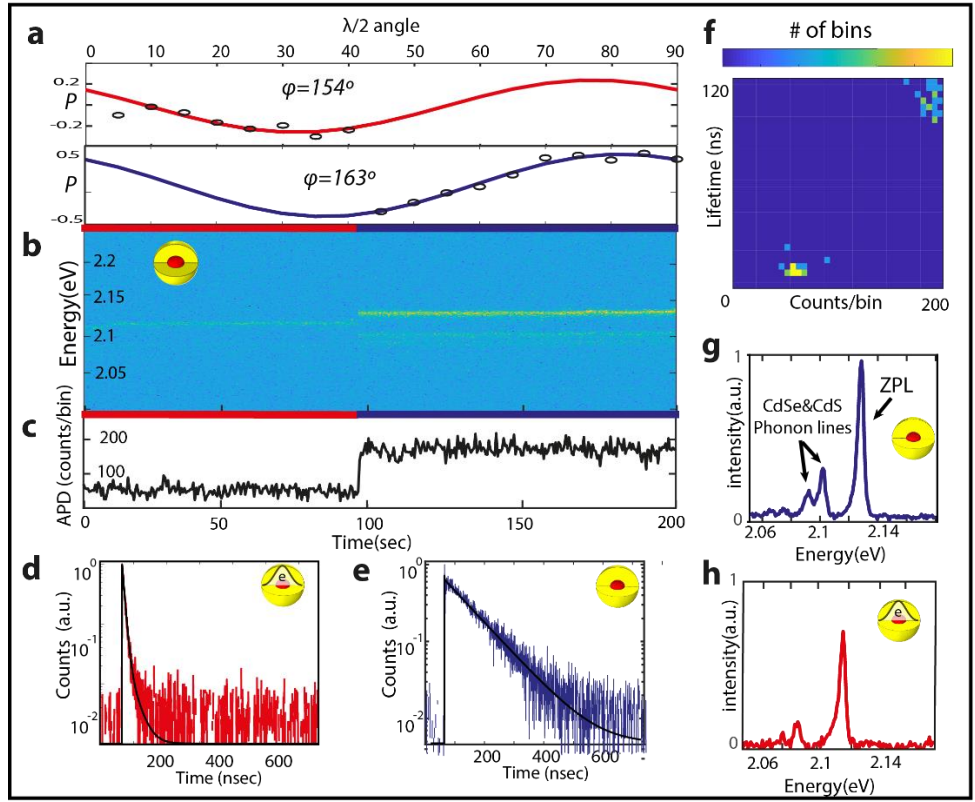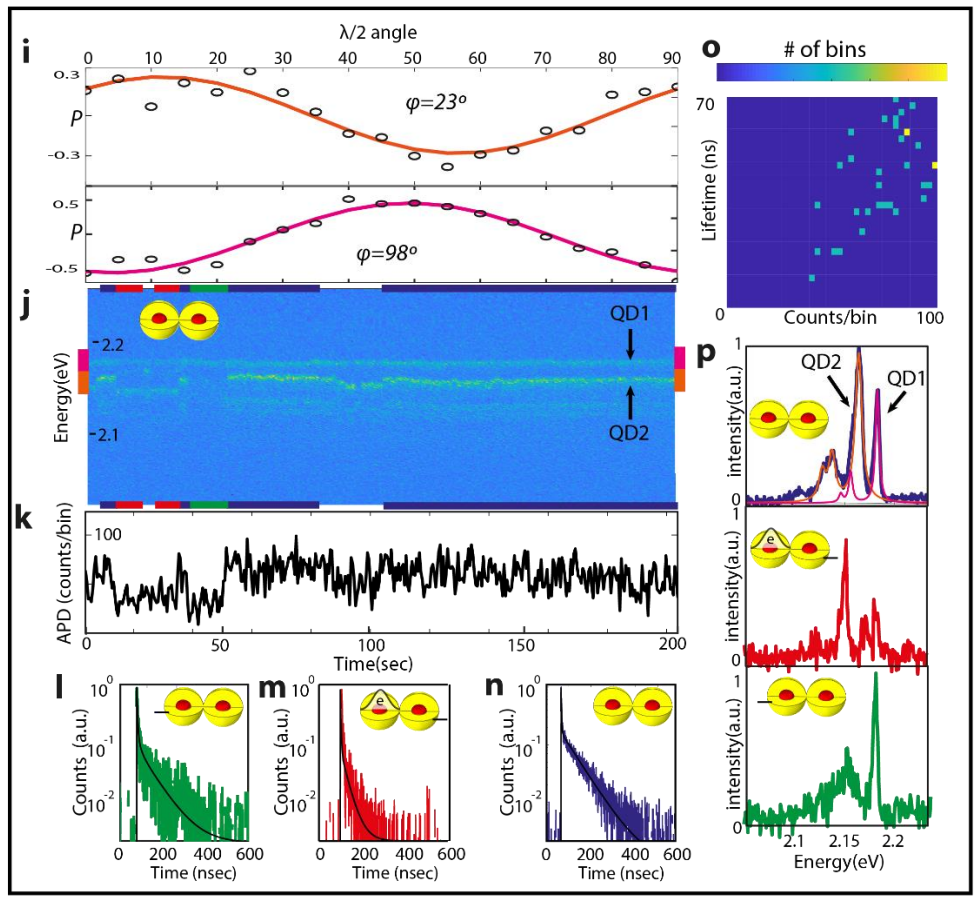

**Figure 2. Single CQD vs. single CQDM spectrum & lifetime data.** (a) Single CQD emission polarization value and angle ($\varphi$) from the red and blue clusters. (b) Time dependent spectrum taken from the EMCCD (V and H replicas merged). (c) Time dependent intensity taken from the APD. Lifetime taken from the counts during the red cluster (d) and from the blue cluster (e). (f) FLID data taken from all counts of the measurement. Summation of all the time dependent spectrums during the blue cluster (g) and the red cluster (h). Insets describing the state of the CQD in each cluster. (i) Single CQDM emission polarization value and angle ($\varphi$) from the magenta and orange spectrum regions. (j) Time dependent spectrum taken from the EMCCD (V and H replicas merged). (k) Time dependent intensity taken from the APD. Lifetime taken from the counts during the green cluster (l), from the red cluster (m) and from the blue cluster (n). (o) FLID data taken from all counts of the measurement. (p) Summation of all the time dependent spectrums during the blue, red and green clusters. Insets describing the state of the CQDM in each cluster.

**Single CQDs & CQDMs spectrum-clusters statistics.**

The origin of multiple emissive peaks as attributed to the two different emitting centers in a CQDM is strengthened by comparison of the emission characteristics of additional single dimers versus single monomers. Following the example of the detailed analysis of a particular single CQDM in comparison with the CQD, we move on to analyze a statistically significant number of single particles of each type utilizing the clustered spectrum analysis described above. Such analysis was done to ~30 single CQDs and ~80 single CQDMs. Since phonon lines maintain the energy difference from the ZPL while it is spectrally diffusing, they cannot explain the extra peaks which are varying without constant energy difference. For each cluster-spectrum we have fitted multiple functions comprised out of a ZPL Lorentzian shaped line and two Lorentzian phonon lines attached to it, red shifted by 27 and 37 meV, corresponding to CdSe and CdS LO phonons, respectively. As an example, figure 2p (blue) shows the fit of two such functions (magenta and orange lines) to the neutral state spectrum. The locations of the ZPLs and their intensities were extracted, together with the number of ZPLs in the cluster-spectrum, the long component of the lifetime when it is present, its amplitude and the polarization angle difference between different peaks in the spectrum.

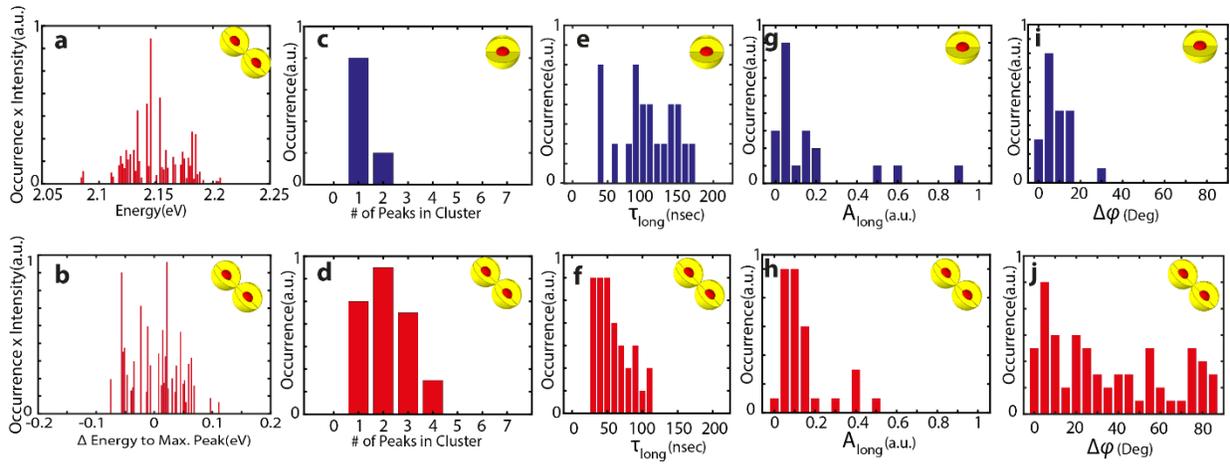

**Figure 3. Single CQDs & CQDMs spectrum clusters statistics.** (a) The occurrence of locations of the ZPLs taken from different CQDM spectrum clusters multiplied by their intensities. (b) The energy distance between different peaks in CQDM cluster spectrums compared to the maximum peak in each cluster. (c) The number of ZPLs in QD spectrum clusters and in CQDM spectrum clusters (d). (e) Occurrence of the long component in the lifetime for QDs and for CQDMs (f). (g) Occurrence of the amplitude of the long component in the lifetime for QDs and for CQDMs (h). Polarization angle difference between different peaks in the spectrum for QDs (i) and for CQDMs (j).

Figure 3a shows the locations of the ZPLs extracted from different CQDMs cluster-spectra multiplied by their intensities. The average ZPLs spectral position for dimers is ~2.15eV and it is ranging between ~2.1-2.2 eV. When we compared the spacing of the other ZPL peaks for dimers relative to the maximum ZPL peak in each cluster (Fig. 3b), these range roughly symmetrically between -70mev to + 70meV. No specific energy spacings emerge. This rules out their assignment to negative or positive trion peaks, as in such cases the spacing from the maximum peak should have been in certain fixed intervals. For example, for a negative trion, the spacing would be in the range of 10-20meV red shift as verified by numerical simulations (Supporting Information fig. S5b).

In addition, we observe that the number of ZPL peaks in the cluster-spectra of CQDMs (Fig. 3d) can reach up to 3 or 4 compared to QDs where it is predominantly only one ZPL peak (Fig. 3c). If the extra dimer peaks were associated with trion emission, in the case of 3 peaks it would mean that the CQDMs were occupying neutral, positively and negatively charged states, all on a time scale faster than the exposure time of the EMCCD (0.5 sec). This is never observed in the case of CQDs and moreover, the case of 4 ZPL peaks cannot be explained by trions alone.

Furthermore, the longest component in the lifetime, associated with each CQDM spectrum-cluster, is ranging between 40-120 nsec (fig. 3 f,h), very similar to the case of monomer CQDs, where it is 40-160 nsec (fig. 3 e,g), and the amplitudes of the long components are also similar. If the extra peaks were trions on a single emitting center of the dimer, the amplitude of the long component should have been much smaller or absent altogether. This has two origins, first - non-radiative Auger relaxation in such case would lead to fast decay. Second – the presence of the extra carrier should lead to bright state occupation at the band edge. [38]

Last but not least, the emission polarization angle difference $\Delta\varphi$ between the different peaks in the CQDM clusters are random (fig. 3j), compared to the case of monomer CQDs where $\Delta\varphi$ is less than 20º even including the phonon lines (fig. 3i). We conclude from all these data comparisons, that the extra peaks in dimers arise from emission from different cores of the CQDM. The different polarization angles are due to different orientations between the two CQDs comprising the CQDM. The energy spread of the ZPL peaks resembles the size dispersion among the monomer CQDs. The number of ZPLs is mostly two. The cases where it is more than two may be attributed to CQD trimers or tetramers. The long lifetime components arise from the neutral state of the CQDMs.

We further note that recent work on single nanoplatelets[28] also shows single particle data with multiple lines attributed to shake up lines from the charged state. In a shakeup process, one of the electrons in the trion takes only part of the energy while the other electron and hole pair recombine radiatively while emitting a slightly red shifted photon. However, a shake-up process can happen only in the weak confinement regime such as in nanoplatelets, but for CQDs and for CQDMs in the strong confinement limit, these transitions are forbidden due to the orthogonality of the electron wave function of the trion and the excited single-electron wave functions.[28,43]

**Observation of simultaneous spectral fluctuations in CQDMs.**

It can already be observed in figure 2j and in many other single CQDMs (see supporting information fig. S1-S4), that simultaneous correlated changes are observed in the emission spectra of the two CQDs comprising the CQDM. In order to explore this phenomenon, we focus on one single CQDM as presented in Fig. 4, in which the emission of the two QDs are in different regions of the spectrums enabling easier tracking of each and every one of the QDs peaks without further ambiguity.

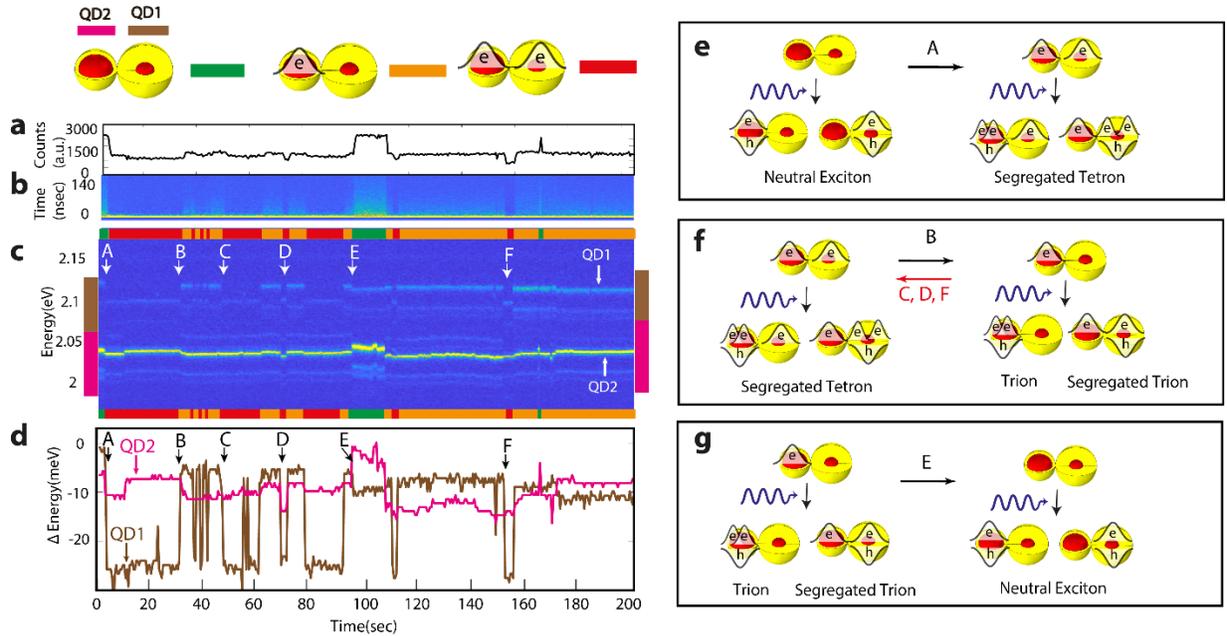

**Figure 4. Simultaneous spectral movements in CQDMs.** (a) The intensity and lifetime data (b) during the measurement. (c) Time dependent spectrum (letters from A-F marks the times when a simultaneous change in the emission spectrum of the two QDs is observed). (d) The maximum peak locations in the brown and magenta regions of the spectrum. Assignment of the states of the CQDM before and after the simultaneous changes in case A (e), in cases B, C, D and F (f) and in case E (g).

Figure 4c shows the time dependent spectrum of this CQDM. We marked with letters from A-F the instances where we observe a simultaneous change in the emission spectra of the two QDs. These simultaneous changes are even seen better in figure 4d, showing the maximum peak locations in the brown (QD1) and magenta (QD2) regions of the spectrum. The intensity (fig. 4a) and lifetime data (fig. 4b) during the measurement teaches us that the CQDM is alternating between three main states during the measurement time. In the green cluster, high intensity is observed, accompanied by a long lifetime component, and this corresponds to the two QDs being neutral. In the orange cluster, there is a component of increased lifetime, intermediate intensity is obtained, and this is accompanied by a ~10meV red shift in the QD2 emission spectrum. This corresponds to a state where only QD2 is charged with another electron. In the third, red cluster, the intensity is lowest, there is only a short component in the lifetime, and there is an 18meV red shift of the QD1 emission compared to its neutral state (the green cluster for example). Here, the two QDs are charged with an additional electron.

With these assignments, the simultaneous changes can be clearly explained. The starting point is of two neutral QDs within the CQDM. Upon excitation, this gives emission from an exciton state in either of the QDs. In case A, the CQDM is changing from this state where the two QDs are neutral, to a state where both the QDs becomes negatively charged. Upon excitation this will give emission from the segregated tetron state (fig. 4e). The segregated tetron is the state where a trion (2 electrons and one hole) is in one of the QDs while another electron is in the other QD. Unlike single QD, where the tetron state involves electron occupation in higher P state, in CQDM the extra electron is occupying the 1S state of the other QD. In case B, the CQDM is changing from a state where both QDs are negatively charged, which upon excitation will give emission from the segregated tetron state, to a state where only QD2 is charged, which upon excitation will give emission either from a trion state in QD2 or from a segregated trion state (fig. 4f). In cases C, D, and F, the CQDM is undergoing the reverse change, from a single charge QD, to a situation where both QDs are charged. In case E, the CQDM is changing from a case where only QD2 is charged, to a case where both QDs are neutral, which upon excitation will give emission from a neutral exciton state (fig. 4g).

The common occurrence for cases B, C, D, E and F is the charging or discharging of one of the QDs while the other QD is not changing its state. However, one can notice that while in cases B, D and E charging or discharging of one of the QDs affects the emission spectrum of the other QD by more than 5meV, in the other cases, C and F, the effect on the other QD is small (shifting by less than 1meV).

Quantitative analysis of the different situations was achieved by solving the self-consistent Schrodinger-Poisson equation using Comsol. In this simulation, the Schrodinger equation for every charge carrier is solved iteratively using the Coulomb potential of all the other charge carriers including its self-potential because of the dielectric mismatch between the nanocrystal and its surrounding. Then, using the solved wave-function, by solving the Poisson equation, a Coulomb potential of the charge carrier is generated. This Coulomb potential is then placed in the Schrodinger equation of the next charge carrier. These equations are solved iteratively for all the charge carriers until they converge. More details are given in in the supporting information (fig. S5 shows how we deduced the dimensions of the CQDM and fig. S6-S8 illustrate the algorithm).

Firstly, the influence of an extra electron in the other QD on the emission of the QD that is not changing its charging state, is addressed. In cases B, C, D and F it is the change between a trion in QD2 to a segregated tetron where two electrons and one hole are in QD2 and one extra electron is in QD1. The shift for QD2 is by less than 1meV. Similarly, a small shift of less than 1meV is calculated also in case E where the change is between a segregated trion where electron and hole are in QD1 and another electron is in QD2, to a neutral exciton in QD1. This therefore leaves the open issue of what might lead to a change of more than 5meV in the emission of the QD, which is not changing its charging state, while the other QD is getting charged?

**Influence of surface charge on the spectrum of CQDMs.**

It is well established in the literature that the blinking phenomenon in the emission of a single nanocrystal is attributed to charging of the QD by the escape of the hole to a surface trap, leaving the QD charged with an extra electron.[39,44] Upon excitation, a trion is formed, which due to competing non-radiative Auger recombination, is a dim state. The nanocrystal is going back to its on state when the hole comes back to the nanocrystal or by an Auger process in which the extra electron is also ejected out of the nanocrystal to a surface trap. Thus, every charging or discharging event is also accompanied with a change of charge on the surface. In order to check what will be the influence of an extra charge on the surface (either positive or negative) on all the emissive states that are present during the measurement, we have added a point charge on the surface of the CQDM in the simulation. Now, the self-consistent Schrodinger-Poisson equations are being solved iteratively, including the Coulomb potential of the surface charge in addition to the Coulomb potential of the other charge carriers.[45]

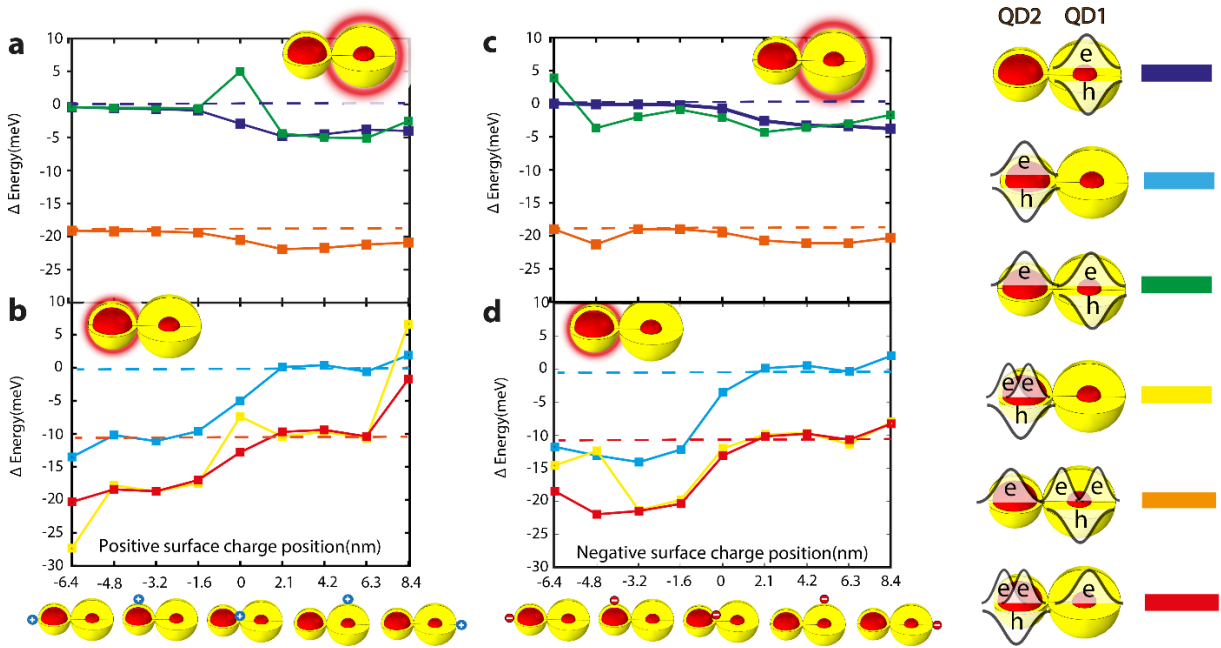

**Figure 5. Influence of surface charge on the spectrum of CQDMs.** The impact of a positive (a) or negative (c) surface charge on the emission spectrum of QD1 in a neutral exciton state (blue), in a segregated trion state (green) and in a segregated tetron state (orange). The impact of a positive (b) or negative (d) surface charge on the emission spectrum of QD2 in a neutral exciton state (light blue), in a trion state (yellow) and in a segregated tetron state (red). In all cases, the dashed line represents the same state but with no surface charge.

Figure 5 shows the impact of a positive (fig. 5a,b) or negative surface charge (fig. 5c,d) on all the states during the measurement, as a function of the location of the surface charge along the long dimer axis, either when the emission comes from QD1 (fig. 5a,c) or from QD2 (fig. 5b,d). Note that we are addressing here the spectral changes. However, the overlap integral, according to our calculation, is changing only by few percent at maximum. This is supported by the stable intensity of the peaks during spectral diffusion (fig 4c). The states without surface charge are designated by a horizontal dashed line. Specifically, the blue dashed lines in fig. 5a,c represent both the neutral and the segregated trion states without surface charge, and the red dashed lines in fig. 5b,d represent both the trion and the segregated tetron states without surface charge.

The energy spacing between two dashed lines in each plot represent the negative trion red-shift compared to the neutral exciton. For example, the spacing between the light-blue to the red dashed lines in figure 5d, represents the red-shifted emission which comes from the QD2 trion state compared to a neutral exciton state in QD2. That is, having an excess electron in QD2 besides the electron-hole pair, and including the case where there is another electron in QD1 that does not cause a major shift. In all cases, we see that when the location of the surface charge is on the QD where the emission comes from, the red-shift, compared to the dashed line which represents the same state without surface charge, is of more than 5meV. When the surface charge is on the other QD, the red-shift is much smaller, on the order of 1meV.

Outlier points with a large shift in other cases, even when the surface charge is on the other QD (e.g. fig. 5b at 8.4nm and fig. 5c at -6.4nm and -4.8nm), are related to hybridization effects. For example, because of the surface charge, the leftover electron wave-function after emission from a trion state is hybridizing, while, when the other charge carriers of the trion state are present, the electron wave-function is localized in one of the QDs. Since the self-potential of hybridized or localized electron wave-function are different, larger energy shifts are observed compared to the case with no surface charge (see supporting information fig. S9).

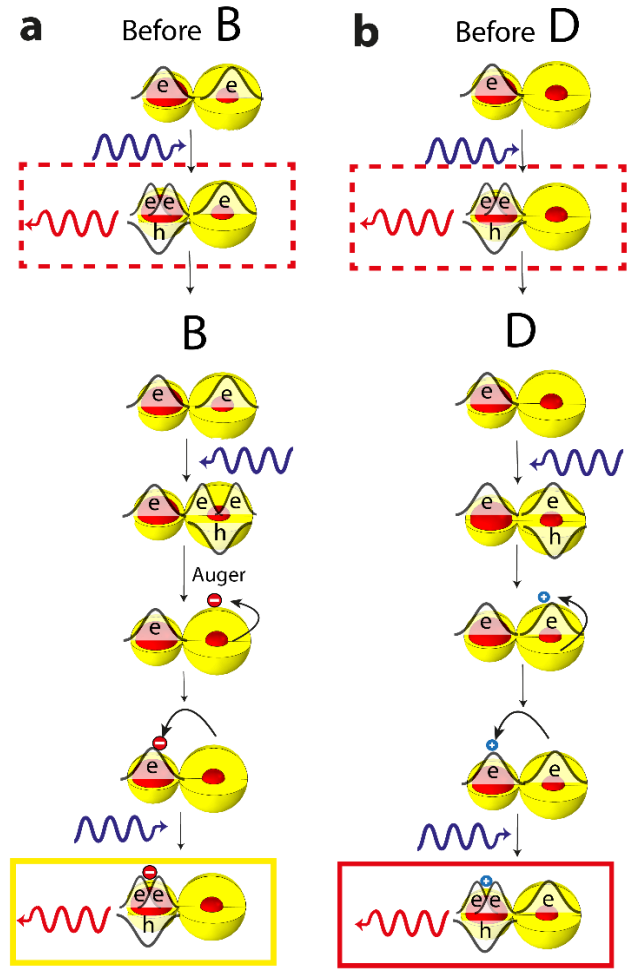

**Figure 6. Assignment of the simultaneous spectrum movements in the CQDM to surface charges.** (a) In case B (b) in case D. Colored rectangles highlighting the emission states before and after the simultaneous spectrum movement and are color coded as the states in figure 5.

Nonetheless, we focus our attention to more plausible situations where the surface charge is not particularly localized in specific regions. This is backed by observation of the larger jumps, by 5meV and more, also for weakly fused and in dimers connected by organic linkers prior to fusion. This has led us to the conjecture that cases B, D and E, where we see that charging or discharging of one QD leads to a red-shifted emission by more than 5meV in the other QD, is a direct observation of a charge movement from the surface of one QD to the surface of the other QD leading to a large red-shift in its emission spectrum. Note that in figure 5 the effect of surface charge can reach up to 15 meV while in the experiment it is more moderate. It is well explained by the screening of the trap while in the simulation the surface charge is taken as a full charge.[46]

More specifically, we next describe in detail such occurrences for particular cases. In case B, before QD1 discharged, both QD1 and QD2 were negatively charged (Fig. 6a). Upon excitation of QD2, the emission from QD2 is coming from a segregated tetron state in which QD2 contains 2 electrons and one hole, while QD1 contains another electron (top red dashed rectangle in fig. 6a). The discharging of QD1 is a result of an excitation of QD1 which led to an Auger process in which one of the electrons gets the energy from another electron-hole pair and is ejected out into a surface trap. This electron moved to the surface of QD2 and led, upon excitation of QD2, to the formation of a trion, along with a negative surface charge (bottom yellow rectangle in fig. 6a). Comparing the two emission states from QD2, before and after event B, now explains the observed large red-shifted emission as predicted by the modelling (the red shift of the yellow line from the red dashed line in fig. 5d).

The same sequence happens also in case E. However, now it is QD2 which is discharging and transferring the negative surface charge to QD1. Comparing the two emission states from QD1, before and after event E (blue dashed line rectangle and blue rectangle in fig. S10b), now explains the observed large red-shifted emission (the red shift of the blue line from the blue dashed line in fig.5c).

In case D, as illustrated in Fig. 6b, the starting point is a negatively charged QD2. QD1 is getting negatively charged upon excitation of an electron-hole pair and ejection of the hole to a surface trap. This positive charge is then transferred to the surface of QD2. Upon excitation of QD2, QD2 contains 2 electrons and one hole plus a positive surface charge, while QD1 contains another electron (bottom red rectangle in fig. 6b). Emission from this state is red-shifted by more than 5meV (fig. 5b the red shift of the red line from the red dashed line) compared to a negative trion in QD2 (top red dashed line rectangle in fig. 6b).

However, cases C and F, where charging of QD1 changes the emission from QD2 by less than 1meV, are cases where the positive surface charge, which was created upon excitation of the neutral QD1 and ejection of the hole to a surface trap, is not transferring to the surface of QD2 (fig. S10a). Thus, the emission of QD2 is hardly changing as predicted by the simulations of such a case (fig. 5b the red line is close to the red dashed line in the right part).

Case A, illustrated in Fig. S10c, is a special case where both the QDs are getting negatively charged simultaneously. A possible explanation to interpret this case is by a starting point where the two QDs are neutral. However, QD2 contains an electron in an inaccessible trap which cannot deliver the electron back into the QD. However, upon charging of QD1 and surface charge transfer of a positive charge to the surface of QD2, the electron on the surface is moving to an accessible trap and by that, charging QD2 as well (fig. S10c).

**Conclusions**

In conclusion, the intricate low temperature emission spectra of CQDMs are measured and analyzed while comparing to the much simpler monomer CQD case. The dimers exhibit a complex spectrum containing multiple peaks that shift with time. Moreover, these peaks show simultaneous changes. Using cryogenic single particle spectroscopy providing simultaneous multivariable information on the emission spectrum, lifetimes, polarization and their temporal fluctuations, we assign the multiple peaks to the emission from either cores of the CQDMs, allowing to directly map the charging state of each of the QDs comprising the CQDM. Utilizing theoretical simulations of the various charging states, we assign the simultaneous spectral changes to electrostatic interactions upon charging or discharging of one of the QDs. These can be electrostatic interactions when one of the QDs is getting charged and thus changing the spectrum of the other QD. In addition, in some instances a larger effect is being observed on one of the QDs when the other is getting charged. This is attributed to surface charge movement from one QD to the other. Lastly, sometimes the two QDs become negatively charged simultaneously. Therefore, we demonstrate that the CQDM architecture, with the two emitting centers sufficiently nearby each other, allows for precise mapping of the charging states and to determine the occurrence and location of surface charges in the system.

The electrostatic interaction between the two QDs composing the dimer presented here, is a step towards quantum information processing applications. For example, CNOT gates based on the charging state of one QD, which changes the spectrum of the other QD. However, as demonstrated here random spectral fluctuation should be avoided in the future, for example, by thicker or graded shells. With the tools and methodology presented herein, other CQDMs could be directly measured and characterized. The modifications in building blocks, surface and synthesis approaches for the CQDMs, can be tested by such methodology where ultimately, we envision achieving highly controlled emission as was demonstrated over the years for the perfection of CQD monomers. Such progress could bring closer the possibility for quantum information applications based on CQDMs as well as to quantum sensing and their incorporation in innovative electro-optic devices.

**Methods**

**Structural characterization.** Transmission electron microscopy (TEM) was performed using a Tecnai G2 Spirit Twin T12 microscope (Thermo Fisher Scientific) operated at 120 kV. High-resolution TEM (HRTEM) measurements were done using a Tecnai F20 G2 microscope (Thermo Fisher Scientific) with an accelerating voltage of 200 kV. High-resolution STEM imaging and elemental mapping was done with Themis Z aberration-corrected STEM (Thermo Fisher Scientific) operated at 300 kV and equipped with HAADF detector for STEM and Super-X EDS detector for high collection efficiency elemental analysis.

**Single particle optical measurements.** Single particle measurements were performed with a home-built microscope (DIY Cerna Components). Dilute solution of QDs in 2 wt% poly(methyl-methacrylate) were spin coated on Silicon substrates leading to minimum separation of 4–5 μm between the dots as confirmed by wide field fluorescence microscopy. The samples were kept in a Helium closed-cycle cryostat (attoDRY800). The excitation light from a pulsed diode laser (EPL475; Edinburgh Instruments) at a repetition rate of 5 MHz/1MHz, was focused through a cold objective (×100; 0.8 NA, Attocube), which was also used for collecting the emission. The emission light was passed through a dichroic mirror (T505lpxr, Chroma) and additional longpass filter (ET542LP) before focusing either onto Avalanche Photodiodes (MPD, 100um SPAD) or a spectrograph (Kymera 328i) with EMCCD (iXon Ultra 888 camera, Andor) through $\frac{\lambda}{2}$ wave-plate and PBS (Thorlabs) mounted on a motorized rotating stage (PRM1Z8,Thorlabs). Time-stamping were performed using multichannel Time Tagger 20 (Swabian Instruments). Spectral time traces and fluorescence lifetimes were extracted from the time-tagged data using home written MATLAB code.

ASSOCIATED CONTENT

**Supporting Information**. Other single CQDMs that simultaneous correlated changes are observed in the emission spectra of the two CQDs comprising the CQDM. The method to deduce the dimensions of the CQDM discussed in fig. 4-6. Illustration of the algorithm used to solve the self-consistent Schrodinger-Poisson equation using Comsol. Explanation of the outlier points in fig. 5 by hybridization effects. Scheme which describe the simultaneous changes in cases A, C, E and F in fig. 4.

AUTHOR INFORMATION

**Corresponding Author**

* Correspondence and requests for materials should be addressed to U.B. (uri.banin@mail.huji.ac.il)

**Present Addresses**

† Present address: The Center for Molecular Imaging and Nuclear Medicine, State Key Laboratory of Radiation Medicine and Protection, School for Radiological and Interdisciplinary Sciences (RAD-X) and Collaborative Innovation Center of Radiological Medicine of Jiangsu Higher Education Institutions, Soochow University, Suzhou 215123, China.

**Author Contributions**

Y.E.P. and U.B. conceived the idea and designed the experiment. Y.E.P built the single particle microscope. J.B.C synthesized the CQDs and CQDMs. Y.E.P wrote the Matlab Codes and analyzed the data. Y.E.P performed the Schrodinger-Poisson self-consistent calculations using Comsol. U.B. and S.K helped in the interpretation of the results. Y.E.P and U.B. co-wrote the manuscript

**Funding Sources**

The research leading to these results has received financial support from the European Research Council (ERC) under the European Union's Horizon 2020 research and innovation programme (grant agreement No [741767], advanced investigator grant CoupledNC). Y.E.P. acknowledges support by the Ministry of Science and Technology & the National Foundation for Applied and Engineering Sciences, Israel. J.B.C and S.K acknowledge the support from the Planning and Budgeting Committee of the higher board of education in Israel through a fellowship. U.B. thanks the Alfred & Erica Larisch memorial chair.

**Competing Interest**

The Authors declare no competing interests.